\documentclass[a4paper, amsfonts, amssymb, amsmath, reprint, showkeys, nofootinbib, twoside]{revtex4-1}
\usepackage[english]{babel}
\usepackage[utf8]{inputenc}
\usepackage[colorinlistoftodos, color=green!40, prependcaption]{todonotes}
\usepackage{graphicx}
\usepackage{dcolumn}
\usepackage{bm}
\usepackage[colorlinks=true,linkcolor=blue, allcolors=blue]{hyperref}
\usepackage{natbib}
\usepackage[T1]{fontenc} 
\usepackage{bbold}
\usepackage{amsmath}
\usepackage{braket}
\usepackage{subfigure}

\begin{document}

\preprint{APS/123-QED}

\title{Fracton-like phases from subsystem symmetries}

\author{J. P. Ibieta-Jimenez}
 \email{pibieta@if.usp.br}
\author{L. N. Queiroz Xavier}
 \email{lucasnix@if.usp.br}
\author{M. Petrucci}
 \email{marzia@if.usp.br}
\author{P. Teotonio-Sobrinho}
 \email{teotonio@if.usp.br}
\affiliation{
 Departamento de F\'isica Matem\'atica, Universidade de S\~ao Paulo, \\Rua do Mat\~ao Travessa R 187, CEP 05508-090, S\~ao Paulo, Brazil.
}

\date{\today}

\begin{abstract}
We study models with fracton-like order based on $\mathbb{Z}_2$ lattice gauge theories with subsystem symmetries in $d=2$ and $d=3$ spatial dimensions. The $3d$ model reduces to the $3$-dimensional Toric Code when subsystem symmetry is broken, giving an example of a subsystem symmetry enriched topological phase (SSET). Although not topologically protected, its ground state degeneracy has as leading contribution a term which grows exponentially with the square of the linear size of the system. Also, there are completely mobile gauge charges living along with immobile fractons. Our method shows that fracton-like phases are also present in more usual lattice gauge theories. We calculate the entanglement entropy $S_A$ of these models in a sub-region $A$ of the lattice and show that it is equal to the logarithm of the ground state degeneracy of a particular restriction of the full model to $A$.

\end{abstract}

\maketitle

\section{Introduction}

Since the discovery of the fractional quantum Hall (FQH) effect \cite{tsui,Laughlin}, it is known that there are quantum phases of matter that cannot be explained by Landau's symmetry breaking theory. Topologically ordered phases, of which FQH states are a standard example, encompass phases that are beyond the scope of Landau's theory. Intrinsic topological order can be characterized as exhibiting, among other properties, a ground state degeneracy that depends on the topology of the underlying space in which the system lives \cite{wen1} and long-range entangled ground states \cite{wen2}, which means that they cannot be transformed into product states by means of local unitary transformations. Another important feature of topological phases is the presence of anyonic excitations, which is essential to the potential application of topological order in fault-tolerant quantum computation \cite{kitaev2,Nayak08}. A classical example of intrinsic topological order is Kitaev's Toric Code (TC) \cite{kitaev2}, first introduced in the context of quantum computation as a quantum error correction code and as a way of implementing a quantum memory. This model can be interpreted as a $\mathbb{Z}_2$ lattice gauge theory, and it is a particular case of a larger class of models known as Quantum Double Models (QDM) \cite{kitaev2, hbombin, buerschaper,ferreira20142d}, which are topologically ordered exactly solvable models based on lattice gauge theories with arbitrary finite gauge groups.

It is also possible to have topological order with short-range entanglement of the ground states if there are global symmetries in the system. Phases with such behavior are known as symmetry-protected topological (SPT) phases, and they are characterized by the fact that entangled ground states cannot be transformed into non-entangled ones by means of local unitary transformations without breaking a global symmetry. The classification of SPT phases is known to be related to the group cohomology of the global symmetry group \cite{Chen-Wen13,kapustin}.

Global symmetries may also coexist with topological order. A topologically ordered system which respects a global symmetry may host anyons which carry fractionalized quantum numbers under this symmetry, a phenomenon known as symmetry fractionalization. For example, global $U(1)$ charge conservation in the fractional quantum Hall effect imposes a fractionalization of the electric charge of some of its quasiparticles \cite{Laughlin}. The effect of global symmetries on topologically ordered states leads to the notion of symmetry-enriched topological (SET) phases. A system is in a symmetry-enriched topological phase if, by breaking the symmetry, the system still presents topological order \cite{mesaros2013classification,lu2016classification,barkeshli2019symmetry,cheng2017exactly}. 

Intrinsic topological order has its low-energy behavior described by topological quantum field theories (TQFTs), which are intimately connected to category theory \cite{turaev}. In particular, string-net models \cite{LevinWen}, a more general framework that captures essential features of topological phases in $d = 2$ spatial dimensions, are obtained from fusion categories. It is becoming clear that, for systems in $d > 2$ dimensions, higher category theory and higher gauge theory \cite{baez2005higher,baez2011invitation} are essential to understand topological order. A variety of models of topological phases obtained from these structures can be found in the literature \cite{bullivant2017topological,delcamp2018gauge,ricardo}.

Recently, it was theorized that a new type of quantum phase of matter lies beyond the topological order framework. The so-called fracton order \cite{Chamon05,Vijay16, Nandkishore18} refers to quantum phases of matter in which distinct ground states cannot be distinguished by local measurements, a feature shared with topologically ordered systems. However, the ground state degeneracy exhibits a subextensive dependence on the system size, in contrast with the constant ground state degeneracy of topologically ordered models. Also, the spectrum of fracton models is composed by quasi-particles with several mobility restrictions, and some of the excitations, the so-called fractons, are completely immobile, i.e., they cannot be moved by string-like operators, in contrast with anyons in topologically ordered models which are free to move around the lattice. 

Based on the mobility of its excitations, it is common to divide gapped fracton phases into two distinct types: in type-I fracton phases, immobile fractons appear at the corners of membrane-like operators, and there are other quasi-particles that can move along sub-dimensional manifolds, for example along lines or planes if the system is $3$-dimensional. In type-II fracton phases, the only excitations are completely immobile fractons and they live at the corners of fractal-like operators. A more detailed description of the different types of fracton phases is presented in \cite{dua2019sorting}. Standard examples of models exhibiting type-I and type-II fracton phases are the X-cube model \cite{Vijay16,weinstein2018absence} and the Haah code \cite{Haah11}, respectively. Both are exactly solvable spin models in $d = 3$ dimensions. Gapped fracton models are related to glassy physics and localization \cite{Chamon05,castelnovo1,castelnovo2, kim2016localization,prem1,pai2019localization} and they present potential applications to quantum information \cite{Haah11,bravyi2011energy,bravyi2011topological,kim20123d,bravyi2013quantum,raussendorf2019computationally, devakul2018universal,stephen2018subsystem}. There are also gapless fracton models arising from the study of symmetric tensor gauge theories \cite{pretko,pretko2017generalized,slagle2018symmetric,bulmash2018higgs,ma2017fracton,pretkofracton}. This approach allows a connection of fracton models with elasticity theory and gravity \cite{pretko2017emergent,pretko2018fracton,gromov2019chiral}, and some gapped fracton phases can be obtained from symmetric tensor gauge theories defined on a lattice by a Higgs mechanism \cite{bulmash2018higgs,ma2018fracton}. The relation between gapless fracton phases and gravity suggests that fracton-like models may be considered as toy models for the holographic principle \cite{yan,yan2019hyperbolic}.

There are ways of generating fracton phases from known topologically ordered models. For example, the X-cube model can be obtained by coupling layers of $2$-dimensional toric codes \cite{ma2017fracton,vijay2017isotropic,slagle2019foliated}. However, the most common manner in which fracton models are obtained is by considering lattice spin systems with subsystem symmetries as generalized Abelian lattice gauge theories \cite{Vijay16}. Fracton phases are then constructed through a process of gauging the subsystem symmetries \cite{slagle1, williamson2016fractal}. Another methods to obtain new fracton phases is by twisting the usual fracton models \cite{song2019twisted} and by enriching $U(1)$ gauge theories with a global symmetry \cite{williamson2019fractonic}.

In this work, we study some models with fracton-like order based on lattice gauge theories with subsystem symmetries in $d = 2$ and $d = 3$ spatial dimensions. By fracton-like order we mean that the excited states of the models considered here are confined to certain regions of the lattice, i.e., they cannot move without some energy cost, and thus they are immobile fractons. However, although the ground state degeneracy of our models exhibits a dependence on the geometry of the lattice, as is the case in most of the standard fracton models, it is not stable under the action of local perturbations, i.e., it is not topologically protected. The $2$-dimensional model we present here is similar to the one introduced in \cite{yan,yan2019hyperbolic}, and serves as a guide to the study of the $3$-dimensional model. This new model reduces to the $3d$ Toric Code when the system is perturbed by operators that break the subsystem symmetry, realising an example of a subsystem symmetry-enriched topological phase. The ground state degeneracy of this new model grows exponentially with the square of the linear size of the system, and there is also a topological contribution to the ground state degeneracy when the system is defined on topologically non-trivial manifolds. Moreover, while this new model has completely immobile fractons as some of its excitations, there are also quasi-particles in the spectrum that are fully mobile. Even though some known models also present ground state degeneracy that grows exponentially with the square of the linear size of the system \cite{petrova2017simple} and mobile charges \cite{bulmash2019gauging, prem2019gauging}, the method introduced here gives an alternative construction of fracton-like models that differs from the usual ones. We calculate the entanglement entropy $S_A$ in a sub-region $A$ of these models and show that $S_A = log(GSD_{\tilde{A}})$, where $GSD_{\tilde{A}}$ is the ground state degeneracy of a particular restriction of the full model to the sub-region $A$, a result in agreement with \cite{ibieta2020topological}. 

The outline of this paper is the following: in Section \ref{sec2}, we review a $2$-dimensional fracton-like model and describe its properties, explaining how its subsystem symmetries are related to its fracton-like properties. In Section \ref{sec3}, we introduce a $3$-dimensional fracton-like model and analyze its fracton properties, borrowing some ideas from Section \ref{sec2}. In Section \ref{ententr}, we calculate the entanglement entropy $S_A$, in a sub-region $A$, of the models studied in the previous sections, and show that they obey the relation $S_A = log(GSD_{\tilde{A}})$, where the meaning of $\tilde{A}$ will be clarified. In Section \ref{sec4}, we make some remarks about how one could study models of fracton-like phases based on gauge theories with arbitrary finite gauge groups.  

\section{Review of fracton-like order in two dimensions}\label{sec2}

In this section, we start by reviewing a simple example of a model exhibiting fracton-like order in two spatial dimensions. This $2d$ model was first introduced in \cite{xu1,xu2} to describe a superconducting state and it is known as the Xu-Moore model or \emph{plaquette Ising model}. The model is shown to have \(1d\) subsystem symmetry in \cite{slagle1,you} and thus considered as a model for fracton-like order in \cite{yan, yan2019hyperbolic}, as we will review. Its classical version is known as the \emph{gonihedric Ising model}, a particular case of the eight-vertex model \cite{yan2019hyperbolic,baxter2016exactly}, studied in the context of string theory and spin-glass physics in \cite{savvidy1994geometrical,savvidy2000system,espriu2004dynamics,espriu2006gonihedric} and references therein.

\subsection{The model}\label{sec:2d-model}

Consider the discretization of a $2$-dimensional oriented manifold $M$. For simplicity, we take the discretization to be described by a square lattice \(K\). The lattice is composed by a set of vertices \(K_0\), a set of links \(K_1\) and plaquettes \(K_2\). To each vertex \(v \in K_0\) we associate a local Hilbert space \(\mathcal{H}_v\) with basis \(\{\ket{1},\ket{-1}\}\). In other words, a spin-$1/2$ degree of freedom sits at each vertex \(v\). Consequently, the total Hilbert space of the model, $\mathcal{H}$, is given by the tensor product of the local Hilbert spaces over all vertices,
\begin{align}
    \mathcal{H} := \bigotimes_v \mathcal{H}_v.
\end{align}

For each plaquette $p \in K_2$, we define the operator 
\begin{align}\label{eq1}
    B_p = \frac{1}{2}\left(\bigotimes_{v \in p}\mathbb{1}_v + \bigotimes_{v\in p}\sigma^z_v\right),
\end{align}
that acts over the spins at the four vertices of $p$. This operator collects the values of spins at the vertices of \(p\), such that it favors configurations with even number of \(\ket{-1}\) around plaquettes. Although this operator seems to be just comparing the degrees of freedom at the vertices around plaquettes, a more physical interpretation of its action will be given in section \ref{sec4}. The global \(\mathbb{Z}_2\) symmetry of the model is made part of the Hamiltonian by means of the projector
\begin{align}\label{eq2}
    A = \frac{1}{2}\left(\bigotimes_{v \in K_0}\mathbb{1}_v + \bigotimes_{v\in K_0}\sigma^x_v\right),
\end{align}
where the tensor product is taken over all vertices in \(K_0\). This operator enforces a global gauge transformation on the system. Given \eqref{eq1} and \eqref{eq2}, the Hamiltonian is defined by:
\begin{align}\label{eq3}
    H = - A - \sum_p B_p.
\end{align}
The global $\mathbb{Z}_2$ operator $X$ given by 
\begin{equation}\label{globalx}
    X = \bigotimes_{v \in K_0}\sigma^x_v
\end{equation}
commutes with $H$.

\subsection{Fracton properties}

There seems to be three essential features that characterize fracton phases of matter: the subextensive behavior of the ground state degeneracy, the fact that ground states are topologically protected and the mobility constraints of the quasi-particles that belong to the spectrum of the model. The quasi-particles that are usually called fractons are completely immobile if considered individually. Bound states of fractons, however, can have increased mobility. Here we show that the model defined in section \ref{sec:2d-model} indeed supports quasi-particles with restricted mobility and a ground state degeneracy that grows exponentially with the system's size. However, this degeneracy can be lifted by local stabilizer operators and thus its ground states are not topologically protected. Hence, when perturbations to the system do not break the subsystem symmetry, we consider this model as an example of fracton-like order.

\subsubsection{Ground State Degeneracy}\label{sec:2d-GSD}
Since the operators $A$ and $B_p$ commute for every plaquette $p$, we can solve this Hamiltonian exactly. Moreover, the operators $A$ and $B_p$ are projectors, so their spectrum is known. This allows us to characterize the ground state subspace of the model as:
\[\mathcal{H}_0 = \{\ket{\psi} \in \mathcal{H}\;|\, A\ket{\psi} = \ket{\psi}\, \text{and }\, B_p\ket{\psi} = \ket{\psi}\},\]
for every plaquette $p \in K_2$. Let us now construct such states. To start, let $\ket{+} \in \mathcal{H}$ be the state where every vertex spin in the lattice is in the \(\ket{+1}_v\) configuration, namely
\[\ket{+}=\bigotimes_{v} \ket{+1}_v.\]
Similarly, the state where all local degrees of freedom are in the \(\ket{-1}_v\) state is written
\[\ket{-}=\bigotimes_{v} \ket{-1}_v.\]
It is not difficult to see that the two states above satisfy the condition \(B_p\ket{\psi}=\ket{\psi}\) for all plaquettes \(p \in K_2\). Then, the state $\ket{\psi_0} = A\ket{+} = A \ket{-} = \frac{1}{2}\left(\ket{+} + \ket{-}\right)$ is a ground state.

\begin{figure}[h!]
    \centering
    \subfigure[]{\label{fig:d-wall-a}\includegraphics[scale= 0.8]{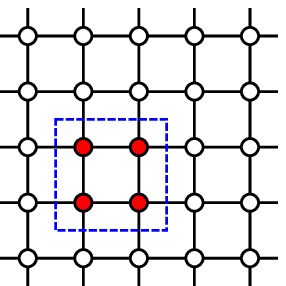}}\qquad\qquad
    \subfigure[]{\label{fig:d-wall-b}\includegraphics[scale=0.8]{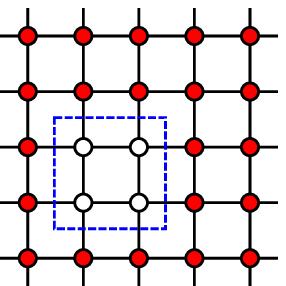}}
    \caption{The domain wall (in blue) of (a) separates two regions with different spin configurations. The same domain wall also determines the separation between configurations of (b). More importantly, the configuration in (a) is gauge equivalent to that of (b).}\label{fig:d-walls}
\end{figure}

\begin{figure}[h]
   \centering
    \subfigure[]{\label{fig:phys-state-1}\includegraphics[scale=0.75]{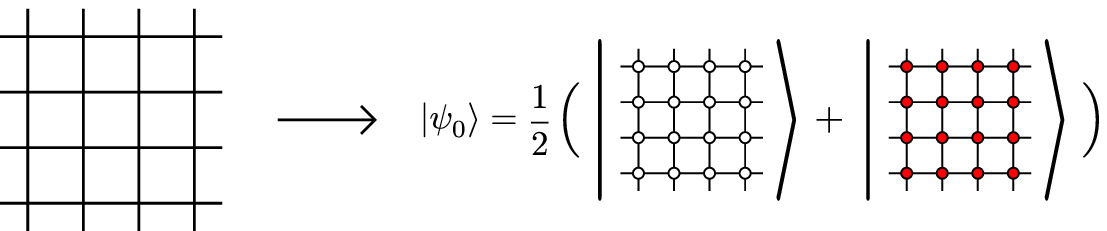}}
    \subfigure[]{\label{fig:phys-state-2}\includegraphics[scale=0.75]{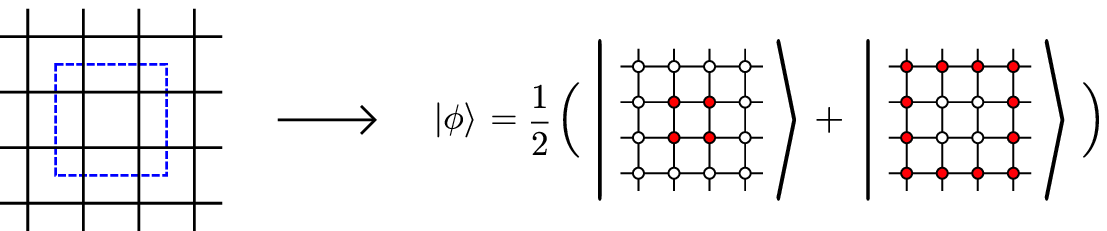}}
    \caption{In (a) the domain wall diagram at the left represents the linear combinations of states at the right, in this case the ground state \(\ket{\psi_0}=A\ket{+}\). In (b) the diagram at the left represents the state $\ket{\phi}$, which in fact is an excited state of the model.}
    \label{fig:phys-states}
\end{figure}

Now, in order to construct other ground states, we will introduce a graphical notation to represent the basis states of \(\mathcal{H}\) as follows; one can color (red) any vertex that holds a \(\ket{-1}_v\) local degree of freedom, see figure \ref{fig:d-walls}. Furthermore, domain walls separating two regions with different spin configuration can be drawn. In general, a single domain wall is associated with two basis states of \(\mathcal{H}\), as shown in figure \ref{fig:d-walls}. However, because of the global gauge transformation the two basis states associated to one domain wall diagram are gauge equivalent. This means that domain walls are enough to represent gauge equivalence classes of basis states, or \emph{physical states}. For example, in figure \ref{fig:phys-states} we show two domain wall diagrams and the respective states they represent.

The trivial ground state, \(\ket{\psi_0}\) is represented by a diagram with no domain walls, as shown in figure \ref{fig:phys-state-1}. On the other hand, the diagram at the left of figure \ref{fig:phys-state-2} stands for a state resulting from a linear combination of states with the given domain wall configuration, this state is actually an elementary excited state of the model as we will see in section \ref{sec:exc-2d}.

\begin{figure}[h]
    \centering
    \subfigure[]{\includegraphics[scale=0.75]{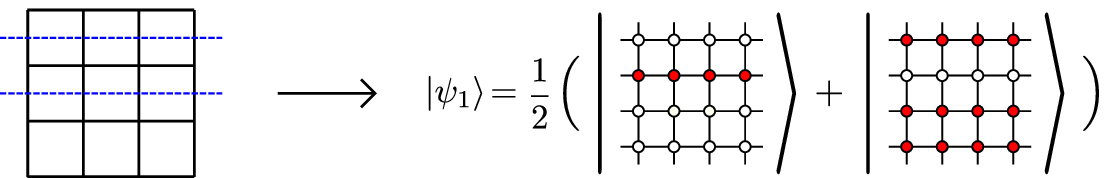}}
    \subfigure[]{\includegraphics[scale=0.75]{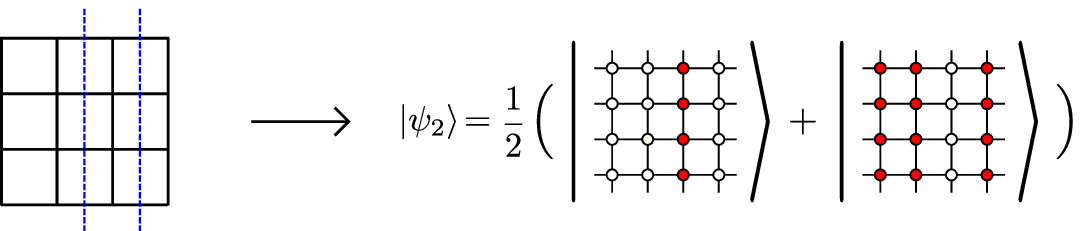}}
    \subfigure[]{\includegraphics[scale=0.75]{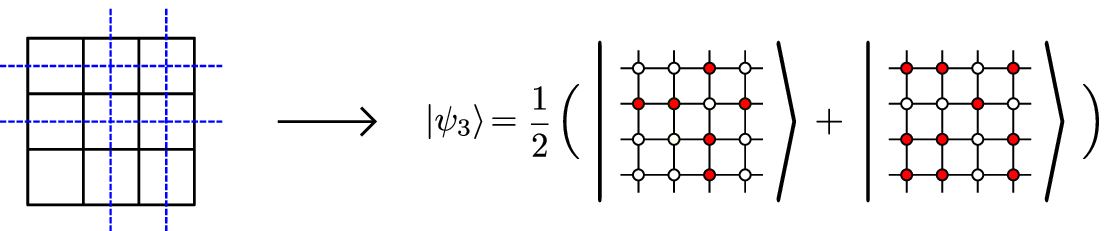}}
    \caption{Some possible ground states of the model. The domain wall lines must begin and end at the boundary of the lattice.}
    \label{fig:gs-2d}
\end{figure}

Other ground states are given by gauge-equivalence classes of states in $\mathcal{H}$ on which $B_p$ acts trivially (i.e., as an identity operator), for every \(p \in K_2\). This means that every state obtained by applying the global gauge transformation on a trivial eigenstate of $B_p$, for all $p$, is a ground state. In order to be invariant under $B_p$, a state must have either an even number of vertices with \(\ket{-1}_v\) spins at each plaquette of the lattice, or no \(\ket{-1}_v\) spins at all. The latter case is taken care of by the state $\ket{\psi_0}$. Examples of the former case are illustrated in figure \ref{fig:gs-2d}. The domain wall lines must begin and end at the boundary of $M$. In case $M$ has no boundary, the starting points of the blue domain wall lines must be identified with its ending points. Essentially, domain wall lines cannot have corners, i.e., every domain wall line that enters a plaquette must exit it in the diametrically opposite side, as opposed to figure \ref{fig:phys-state-2}, which clearly represents an excited state because it has plaquettes with an odd number of vertices with spin $\ket{-1}$. Since the global gauge transformation $A$ does not change domain wall diagrams, any domain wall configuration that represents a (gauge equivalent) linear combination of trivial eigenstates of $B_p$, for every $p$, is a ground state.  

Note that we can have ground states with an arbitrary number of domain wall lines in both directions. If $M$ is a manifold with boundary and has dimension $L_x \times L_y$, this means that we can construct $2^{L_x}$ states with domain wall lines in the \(x-\)direction and $2^{L_y}$ states with domain wall lines in the \(y-\)direction, giving a total of
\begin{equation}\label{eq4}
    GSD = 2^{L_x + L_y}
\end{equation}

possible ground states. This shows the subextensive behavior of the ground state degeneracy, which is characteristic of fracton models. 

Nevertheless, there are local operators that commute with $A$ and $B_p$, for every $p \in K_2$, that can be added to the Hamiltonian (\ref{eq3}) which may destroy this degeneracy. As we will see in more detail in section \ref{sub4.1}, we can define, for every link $l \in K_1$ in the lattice, the $0$-holonomy operator \cite{ricardo,ibieta2020topological}
\begin{equation}
    B_l = \frac{1}{2}\left(\bigotimes_{v \in \partial l}\mathbb{1}_v + \bigotimes_{v \in \partial l} \sigma^z_v\right).
\end{equation}
For each plaquette $p \in K_2$, $B_p$ can be regarded as an operator that compares the $0$-holonomy of parallel links that belong to the boundary $\partial p$ of $p$. By this we mean that $B_p$ gives an eigenvalue equal to one whenever parallel links in $p$ have the same value of $0$-holonomy, and zero otherwise. Fix a plaquette $p' \in K_2$ and subtract of the Hamiltonian (\ref{eq3}) a $0$-holonomy operator $B_{l'}$, where $l'$ is any link in the boundary $\partial p'$ of $p'$. The new Hamiltonian is given by
\begin{eqnarray}
    H' = - A - \sum_{p \neq p'}B_p - B_{p'} - B_{l'}.    
\end{eqnarray}
The ground states of $H'$ must have a $0$-holonomy value of one for the link $l'$, which means that the spins at the vertices in the boundary of $l'$ must be aligned. This introduces an additional constraint to the number of possible ground state configurations of the plaquette $p'$, and thus it reduces the ground state degeneracy. Now, if we subtract two $0$-holonomy operators, $B_{l_1}$ and $B_{l_2}$, for two arbitrary links $l_1$ and $l_2$ in the boundary $\partial p'$ of $p'$, this gives the following new Hamiltonian
\begin{eqnarray}
    H'' = - A - \sum_{p \neq p'}B_p - B_{p'} - B_{l_1} - B_{l_2},   
\end{eqnarray}
for $l_1, l_2 \in \partial p'$. The ground state of the system now must have a $0$-holonomy value equal to one for both links $l_1$ and $l_2$ in $\partial p'$. This means that the spins at the vertices of each link in question must be aligned. If $l_1$ and $l_2$ are parallel to each other, this implies that the configuration of the plaquette $p'$ is fixed; it either has all spins up or all spins down, and both are related by the global gauge transformation $A$, i.e., they represent the same physical state. An identical situation happens if $l_1$ and $l_2$ are perpendicular to each other. Therefore, adding $0$-holonomy operators for the plaquette $p'$ fixes its state, reducing the number of possible ground states. One can immediately see that, if we were to do the same process for every plaquette in the lattice, the degeneracy would be destroyed. Thus, we can decrease the ground state degeneracy shown in equation (\ref{eq4}) by adding local $0$-holonomy operators, and therefore the ground states are not topologically protected. We can move from one state in the ground state subspace to another by applying combinations of $0$-holonomy operators.

This discussion can be summarized by noting that the model has subsystem symmetries given by operators that flip all spins along a straight line in the lattice. The ground state degeneracy (\ref{eq4}) can be calculated by counting the number of such operators, and the $0$-holonomy operators explicitly break this subsystem symmetry, thus drastically reducing the number of ground states. It follows that, when the subsystem symmetry is respected, i.e., when perturbations don't break this symmetry, the model presents fracton-like order. 

\subsubsection{Fracton excitations}\label{sec:exc-2d}

The excited states of the model $\ket{\phi} \in \mathcal{H}$ are states for which either $A\ket{\phi} = 0$ or, for some plaquette $p \in K_2$, $B_p\ket{\phi} = 0$. The excited state coming from the condition on the $A$ operator is usually called \emph{charge}. It is created by acting locally with $\sigma^z$ on a single (arbitrary) vertex over a ground state of the model. The global nature of the gauge transformation makes it impossible to localize the charge, since we can only know whether a charge is present or not. For this reason, the charge is said to be global.

\begin{figure*}
    \centering
    \includegraphics[scale=0.9]{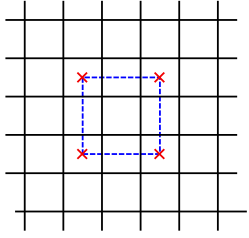}
    \caption{The configuration of figure \ref{fig:phys-state-2} has four fractons, represented here as four red crosses.}
    \label{fig:ee-2d}
\end{figure*}

Plaquette excitations live at plaquettes that have a spin configuration with an odd number of vertices with spin $\ket{-1}_v$. Therefore, they live at the corners of domain walls. For example, the configuration in figure \ref{fig:phys-state-2} has four excitations living at the four corners of the domain wall, as explicitly shown in figure \ref{fig:ee-2d}. We can move pairs of excitations along straight lines, but individual excitations cannot be moved without costing energy to the system, and so they are essentially immobile. Therefore, plaquette excitations in this model are completely immobile fractons, and indeed the system described by the Hamiltonian in equation (\ref{eq3}) exhibits fracton-like order in two dimensions. The arguments made for the calculation of the fracton properties of this model will be important to the study of other models we will define in the following sections.  

\section{Fracton-like order in three dimensions}\label{sec3}

Here we introduce a model of fracton-like order in three spatial dimensions which reduces to the $3d$ Toric Code when subsystem symmetry is broken. This model exhibits some uncommon features, usually not present in the standard examples of fracton phases found in the literature. It is based on a $\mathbb{Z}_2$ lattice gauge theory with slightly modified holonomy operators, as we show in the following subsection.

\subsection{The model}\label{sec:3d-model}

Let's consider a $3$-dimensional manifold $M$ discretized by a regular cubic lattice. At each link $l$, we have a spin-$1/2$ degree of freedom, and the total Hilbert space of the model, which we call $\mathcal{H}$, is a product of all Hilbert spaces that sit at every link of the lattice. For each vertex $v$, we define the local gauge transformation which acts over spins at each link that shares the vertex $v$ as follows:
\begin{equation}\label{eq:Av-3d1}
    A_v \;\vcenter{\hbox{\includegraphics[scale=0.64]{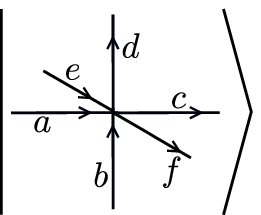}}}\, = \frac{1}{2}\left(\;\vcenter{\hbox{\includegraphics[scale=0.64]{av-ket1.eps}}}\, + \;\vcenter{\hbox{\includegraphics[scale=0.64]{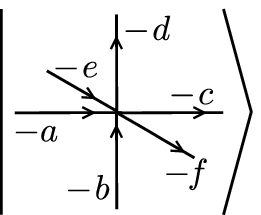}}}\,\right),
\end{equation}
where \(a,b,c,d,e,f \in \{1,-1\}\).

Next, for each elementary cube $c$, we define three holonomy operators, $B_c^{(x)}, B_c^{(y)}$ and $B_c^{(z)}$. To write them in a neat way, we represent the action of $\sigma^z$ operators by coloring links, that is, links in blue are the ones over which a $\sigma^z$ operator act.

\begin{eqnarray}\label{eq:Bc-3d}
    B_c^{(x)} \;\vcenter{\hbox{\includegraphics[scale=0.57]{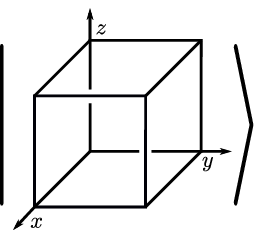}}}\, = & \dfrac{1}{2} \left( \;\vcenter{\hbox{\includegraphics[scale=0.57]{Bc-ket1.eps}}}\,+ \;\vcenter{\hbox{\includegraphics[scale=0.57]{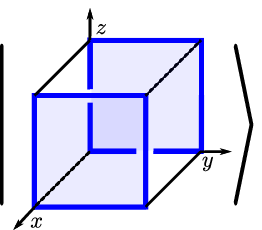}}}\,\right), \label{eq:Bc-3dx}
\end{eqnarray}
\begin{eqnarray}
    B_c^{(y)} \;\vcenter{\hbox{\includegraphics[scale=0.57]{Bc-ket1.eps}}}\, = & \dfrac{1}{2} \left( \;\vcenter{\hbox{\includegraphics[scale=0.57]{Bc-ket1.eps}}}\,+ \;\vcenter{\hbox{\includegraphics[scale=0.57]{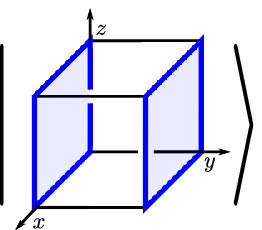}}}\,\right), \label{eq:Bc-3dy}
\end{eqnarray}
\begin{eqnarray}
    B_c^{(z)} \;\vcenter{\hbox{\includegraphics[scale=0.57]{Bc-ket1.eps}}}\, = & \dfrac{1}{2} \left( \;\vcenter{\hbox{\includegraphics[scale=0.57]{Bc-ket1.eps}}}\,+ \;\vcenter{\hbox{\includegraphics[scale=0.57]{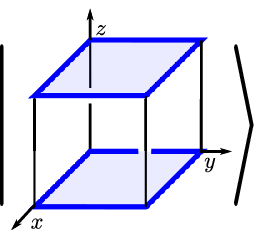}}}\,\right). \label{eq:Bc-3dz}
\end{eqnarray}

The operator $B^{(\mu)}_c$, for each direction $\mu = x, y, z$, checks if the holonomies of two opposite plaquettes in the direction $\mu$ are equal. This is obtained by taking the product of holonomies of the two plaquettes in the boundary of $c$ whose surfaces are orthogonal to $\mu$. If both plaquettes have the same holonomy, this product is equal to $+1$ and we have an eigenstate of $B_c^{(\mu)}$ with eigenvalue $+1$. Likewise, if the two opposing plaquettes have different holonomies, the product is equal to $-1$ and we have an eigenstate of $B_c^{(\mu)}$ with zero eigenvalue, an excited state. We will say that a state has trivial holonomy in the direction $\mu$ if it is invariant under $B_c^{(\mu)}$, for every cube $c$ in the lattice. The Hamiltonian is then given by
\begin{align}\label{eq:H-3D-Z2}
     H = -\sum_vA_v - \sum_c \left(B^{(x)}_c + B^{(y)}_c + B^{(z)}_c\right).
 \end{align} \raggedbottom

\subsection{Fracton properties}

\subsubsection{Ground State Degeneracy}\label{sec:gsd3d}

\begin{figure}[h]
    \centering
    \includegraphics[scale=1]{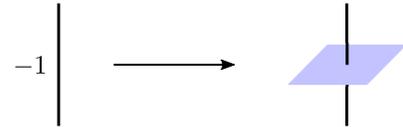}
    \caption{Links holding a \(\ket{-1}_l\) spin are represented by being crossed by a blue dual surface.}
    \label{fig:6}
\end{figure}

As in the $2d$ model, the operators $A_v$ and $B^{(\mu)}_c$ commute for all vertices $v$, cubes $c$ and directions $\mu$ in the lattice, so they can be diagonalized simultaneously. Also, the operators defined in equations \eqref{eq:Av-3d1}, \eqref{eq:Bc-3dx}, \eqref{eq:Bc-3dy} and \eqref{eq:Bc-3dz} are all projectors. This implies that the ground state of the model is given by all gauge-equivalence classes of states with trivial holonomy in all directions. Therefore, we must search for states such that, at each cube of the lattice, opposite plaquettes at each direction have the same holonomy. A natural ground state is $\ket{\psi_0} = \prod_v A_v\ket{+}$, where $\ket{+} \in \mathcal{H}$ is the state of the system where every link is in the \(\ket{+1}_l\) state. To visualize these states, we introduce a graphical notation as follows: whenever a link has spin \(\ket{-1}_l\), we draw a blue dual plaquette, as shown in figure \ref{fig:6}, while links with spin \(\ket{+1}_l\) have no additional drawings.

In this graphical notation, the action of $A_v$, for some vertex $v$, over the state $\ket{+}$ is understood as introducing a blue closed surface around the vertex $v$. That is,
\begin{widetext}
\begin{align}\label{eq:Av-3d}
    A_v \;\;\vcenter{\hbox{\includegraphics[scale=0.40]{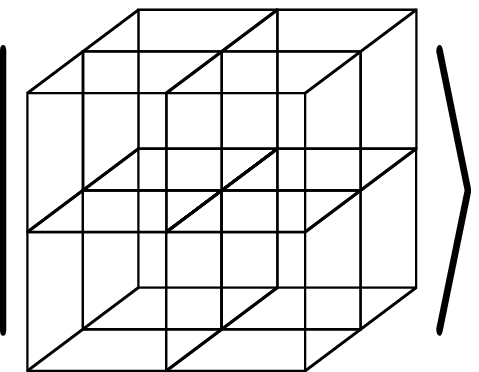}}}\, = \dfrac{1}{2} \left(\; \;\vcenter{\hbox{\includegraphics[scale=0.40]{av-3d-ket1.eps}}}\, +\;\vcenter{\hbox{\includegraphics[scale=0.40]{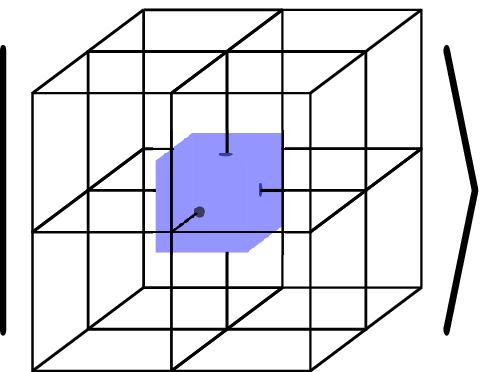}}}\;\; \right), 
\end{align}
\end{widetext}
where the vertex \(v\) is the one at the center of the cubic lattice.
Therefore, the state $\ket{\psi_0}$ is the superposition of all closed surfaces one can draw around vertices in the lattice and this ground state, as shown in equation (\ref{eq:gs-3d}), can be interpreted as a membrane gas much like the loop gas ground state of the toric code.

\begin{figure}[h!]
    \centering
    \subfigure[]{\includegraphics[scale=0.9]{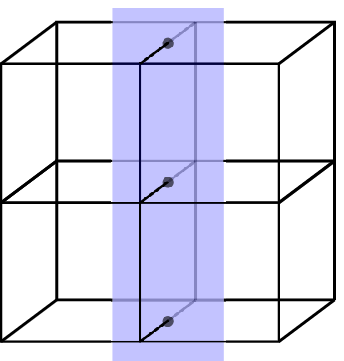}}\qquad \qquad 
    \subfigure[]{\includegraphics[scale=0.9]{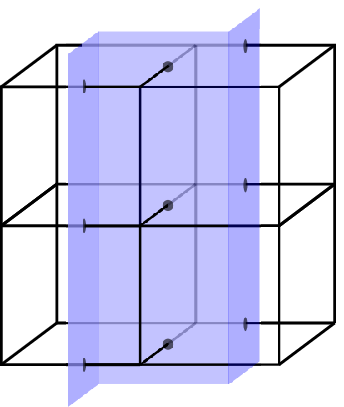}}
    \caption{We show two possible ground state configurations of the Hamiltonian in \eqref{eq:H-3D-Z2}. The membrane must begin at the boundary of $M$ and end at the diametrically opposite region. This means that, inside $M$, the membrane cannot curve to perpendicular directions.
    }
    \label{fig:9}
\end{figure}

\begin{figure*}
    \centering
    \subfigure[]{\label{fig:ee1-3d}\includegraphics[scale=0.8]{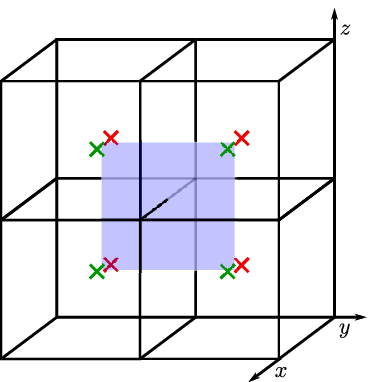}}\qquad \qquad 
    \subfigure[]{\label{fig:ee2-3d}\includegraphics[scale=0.8]{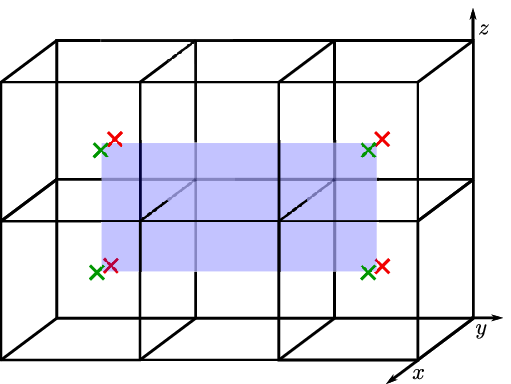}} \\
    \subfigure[]{\label{fig:ee3-3d}\includegraphics[scale=0.8]{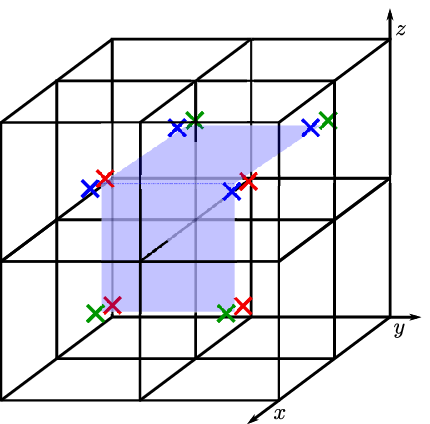}}
    
    \caption{(a) An state represented by the membrane diagram inside $M$ has flux excitations and the four cubes at its corners. The red crosses represent what we call $z$-fluxes, excitations of $B_c^{(z)}$. Whereas the green ones are $y$-fluxes, excitations of $B_c^{(y)}$. Blue crosses stand for \(x\)-fluxes, excitations of \(B_c^{(x)}\) (b) Excitations can freely move as long as the number of corners remains invariant. (c) If a membrane bends into a orthogonal direction, flux excitations are created at every corner it has.}
    \label{fig:11}
\end{figure*}

In the case that the manifold \(M\) has the topology of a \(3\)-torus,  non-contractible closed surfaces give different equivalence classes of ground states. This increases the \(GSD\) with topological terms coming from the $3d$ Toric Code. In other words, the ground states of the $3d$ Toric Code are also ground states of our model, and there is a purely topological contribution to the ground state of the Hamiltonian \eqref{eq:H-3D-Z2}. However, we are more interested in the contribution to the ground state of $H$ that grows exponentially with the system size, the \emph{subextensive terms}. For this reason we consider \(M\) as having the topology of a 3-dimensional ball with dimensions $L_x \times L_y \times L_z$. States represented by membranes beginning at one of the boundaries of $M$ and ending at the diametrically opposite boundary are also ground states of the model. For instance, the state represented by figure \ref{fig:9}(a). Membranes can have arbitrary shapes in every one of the three directions as long as they end at the boundaries of \(M\), as in figure \ref{fig:9}(b). If the membranes do not end at the boundaries of $M$, we have an excited state, as in figure \ref{fig:ee1-3d}, where we have a link with spin $\ket{-1}$ shared by four cubes, which yields an excited state of cube operators in the $z$ and $y$ directions. In the interior of $M$, membranes cannot bend to perpendicular directions, for if they do we get excitations of cube operators at the folding regions of the bent membranes, as in figure \ref{fig:ee2-3d} where the membrane of figure \ref{fig:ee1-3d} is folded into the $x$ direction, giving rise to excitations of $B_c^{(x)}$ operators at the folding line. The gauge transformation acts at vertices and can be pictured as adding a closed (dual) surface around the vertex it acts, see eq. \eqref{eq:Av-3d}. Thus, gauge transformations can only deform membranes without changing their boundary.

\begin{widetext}
\begin{align}\label{eq:gs-3d}
    \ket{\psi_0} = N\, \left(\;\vcenter{\hbox{\includegraphics[scale=0.4]{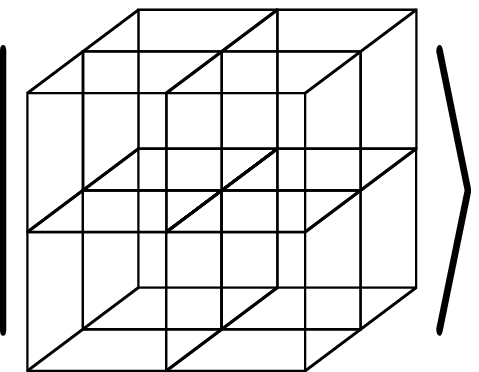}}}\, + \;\vcenter{\hbox{\includegraphics[scale=0.4]{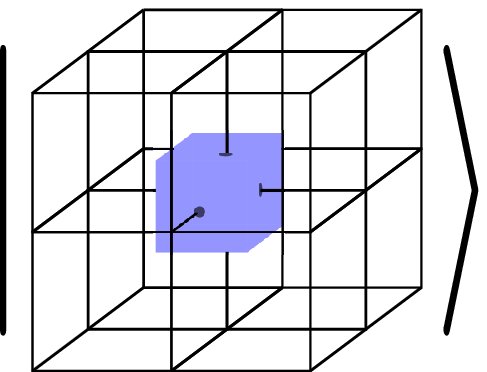}}}\, +\;\vcenter{\hbox{\includegraphics[scale=0.4]{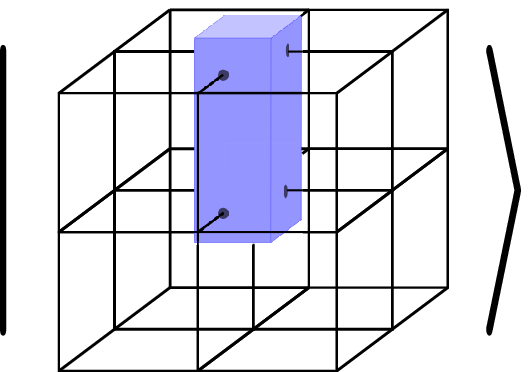}}}\, + \;\vcenter{\hbox{\includegraphics[scale=0.4]{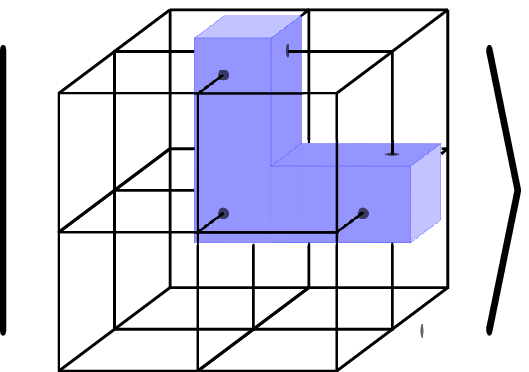}}}\, + \dots \right),
\end{align}
\end{widetext}

\begin{figure}[h]
    \centering
    \includegraphics[scale=0.8]{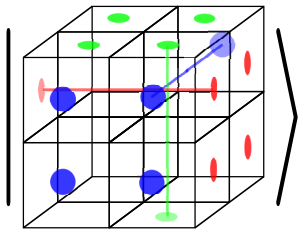}
    \caption{To count the number of ground states of the Hamiltonian (\ref{eq:H-3D-Z2}), we draw straight lines in the interior of $M$ ending at dots on its boundary $\partial M$. These straight lines are boundary lines of the membranes representing ground states (see figure \ref{fig:9}). The problem then reduces to counting how many dots we can draw on the boundary of $M$.}
    \label{fig:10}
\end{figure}

We can have arbitrary compositions of such membrane configurations in every direction. Since the boundary lines of the membranes are gauge-invariant, we use them to count how many possible ground state configurations we can construct in this model. Therefore, the problem of counting ground states reduces to the problem of counting how many straight lines, beginning at one side of the boundary of $M$ and ending at the diametrically opposite side, can be drawn on the manifold. We represent as a dot in the boundary of the manifold the beginning of a line that extends, in a straight fashion, through the interior of $M$ to the diametrically opposite boundary point, as in figure \ref{fig:10}. Each plaquette in the boundary of $M$ either has a dot on it or it doesn't. For each boundary plane of $M$, there can be $2^{N_p}$ configurations of plaquettes with dots, where $N_p$ is the number of plaquettes on the plane in question. Since $M$ has dimensions $L_x \times L_y \times L_z$, the number of plaquettes in the boundary plane with dimensions $L_i \times L_j$ is $L_iL_j$, where $i,j = x,y,z$ and $i \neq j$. Thus, there are 
\begin{equation}\label{eq10}
    GSD = 2^{L_xL_y + L_xL_z + L_yL_z}
\end{equation}
possible ground states. It is useful to think of the ground states of this model as condensations of the the $3d$ Toric Code model. Note that every ground state of the $3d$ Toric Code is a ground state of our model. Moreover, some excited states of the $3d$ Toric Code are ground states of our model. In particular, the flux excitations of the TC that lie on a single plane are ground states of our model as well.

However, as happened to the $2d$ model of section \ref{sec2}, here there are local operators that commute with the Hamiltonian (\ref{eq:H-3D-Z2}), i.e., local symmetry operators, which can lift the degeneracy given by equation (\ref{eq10}) to that of the $3d$ Toric Code. To see this, fix a cube $c'$ in the lattice and subtract from the Hamiltonian (\ref{eq:H-3D-Z2}) a $3d$ Toric Code plaquette operator $B_{p'}$, where $p'$ is some arbitrary plaquette in the boundary $\partial c'$ of $c'$. We have the new Hamiltonian
\begin{equation}
    H' = -\sum_v A_v - \sum_{c, \mu}B_c^{(\mu)} - B_{p'},
\end{equation}
where $\mu = x,y,z$. Ground states of $H'$ must have a $1$-holonomy value of one for the plaquette $p'$. Since the cube operators $B_c^{(\mu)}$ constrain parallel plaquettes to have the same $1$-holonomy in the ground state, the plaquette which is parallel to $p'$ will also have a $1$-holonomy value of one. This reduces the number of ground state configurations the cube $c'$ can have, thus reducing the ground state degeneracy. Now, if we subtract from the Hamiltonian (\ref{eq:H-3D-Z2}) $3d$ Toric Code plaquette operators $B_{p'}$ for four of the six plaquettes in the boundary $\partial c'$ of $c'$, the new Hamiltonian is then given by
\begin{equation}
    H'' = -\sum_v A_v - \sum_{c, \mu}B_c^{(\mu)} - \sum_{p' \in \partial c'}B_{p'},
\end{equation}
where $\mu = x, y, z$ and $B_{p'}$ are the $1$-holonomy operators of the $3d$ Toric Code for four specific plaquettes $p' \in \partial c'$. Now, the ground state of the model must have the four chosen plaquettes with holonomy equal to one. This fixes the allowed ground state configurations of the whole cube $c'$, reducing further the number of allowed ground states. Subtracting $3d$ TC plaquette operators for every plaquette in the lattice, the ground state degeneracy would end up being that of the $3d$ TC, because the contribution given by equation (\ref{eq10}) would be destroyed and only the topological terms would survive. So, this model is not stable under local perturbations, reducing to the $3d$ Toric Code when local operators are added, and therefore it gives an example of a subsystem symmetry enriched topological phase, showing that not only topological phases enriched by global symmetries hosts fractonic behavior \cite{williamson2019fractonic}, but also subsystem symmetry-enriched ones. The question of whether the model presented here can be protected by a global symmetry is an open one. 

\begin{figure}[h!]
    \centering
    \subfigure[]{\label{fig:TC_excited}\includegraphics[scale= 0.8]{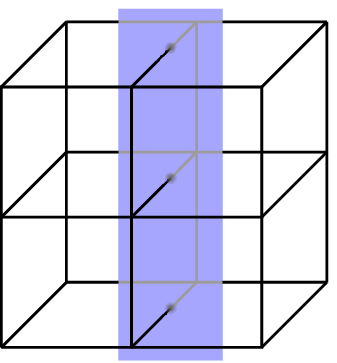}}\qquad\qquad
    \subfigure[]{\label{fig:TC_subst}\includegraphics[scale=0.8]{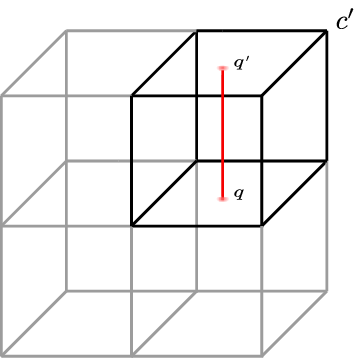}}\qquad\qquad 
    \subfigure[]{\label{fig:TC_excited2}\includegraphics[scale=0.8]{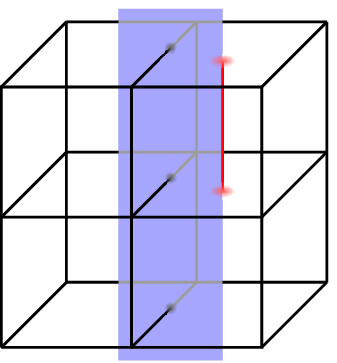}}\qquad\qquad 
    \subfigure[]{\label{fig:TC_excited3}\includegraphics[scale=0.8]{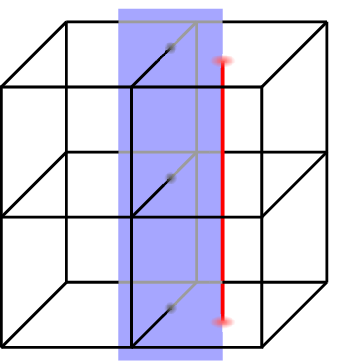}}
    \qquad\qquad 
    \subfigure[]{\label{fig:TC_excited4}\includegraphics[scale=0.8]{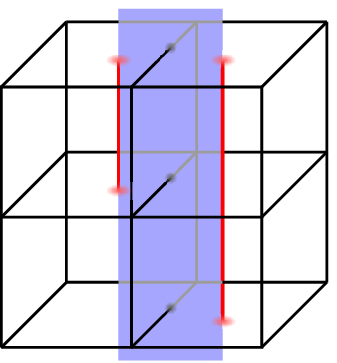}}
    \qquad\qquad 
    \subfigure[]{\label{fig:TC_excited5}\includegraphics[scale=0.8]{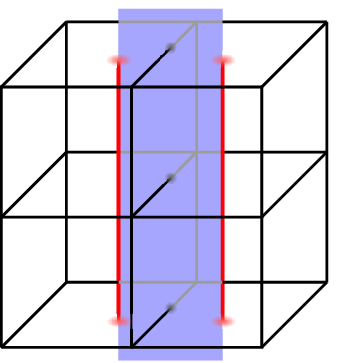}}
    \caption{In (a), we show an excited state of the Toric Code. In (b), we illustrate the graphical notation corresponding to add a cube operator $B_{c'}^{(z)}$ and remove the plaquette operators $B_q$ and $B_{q'}$ of the Toric Code Hamiltonian. In (c), we take the model defined by the Hamiltonian obtained in figure (b) and consider the TC excited state of figure (a) in this new context. It is still an excited state, but its energy is reduced by two unities. In (d), by continuously extending the red line of figure (c), we obtain a state with an even more reduced energy. In (e) and (f), the same procedure is done for all cubes hosting a half of the ribbon, resulting in a ground state.}
    \label{fig:replacing_ops}
\end{figure}

We saw that we can go from the fracton-like model defined by equation \eqref{eq:H-3D-Z2} to the Toric Code by adding plaquette operators that break the subsystem symmetry. A reasonable question is then whether there is a way to go from the $3d$ Toric Code to the fracton-like model. To answer it, suppose we start with the $3d$ Toric Code defined on the $3$-torus, whose Hamiltonian is \[H_{TC} = -\sum_v A_v - \sum_p B_p.\] In the graphical notation introduced in this section, ground states of the Toric Code are represented by closed dual surfaces. Dual surfaces with boundary correspond to plaquette excitations. Consider an excited state of the Toric Code represented by a non-contractible dual ribbon, as in figure \ref{fig:TC_excited}. In this figure, there are flux quasi-particles at all plaquettes along the $z$ direction. The boundary of this ribbon is composed by two non-contractible curves. The energy of this state is two times the length $L_z$ of the torus in the $z$ direction. 

We will now replace plaquette operators of the Toric Code with the cube operators defined in equations \eqref{eq:Bc-3dx}, \eqref{eq:Bc-3dy} and \eqref{eq:Bc-3dz}. To better visualize what is happening, we assign the following graphical notation to the process of replacing plaquette with cube operators: we draw a straight red line connecting two parallel plaquettes whose operators are removed from the TC Hamiltonian. Then, to any cube hosting such a red line we associate a cube operator $B_c^{(\mu)}$, where $\mu$ is the direction parallel to the red line. As an example, in figure \ref{fig:TC_subst}, we perform this procedure to the cube $c'$, associating to it an $B_{c'}^{(z)}$ operator. In this example, the resulting Hamiltonian is \[H'_{TC} = -\sum_v A_v - \sum_{p\neq q,q'}B_p - B^{(z)}_{c'}.\]

Then, consider again the Toric Code excited state of figure \ref{fig:TC_excited}, but now with a red line linking two parallel plaquettes in the $z$ direction, as in figure \ref{fig:TC_excited2}. Now, the Hamiltonian of the model has an $B_c^{(z)}$ operator corresponding to the cube hosting the red line. Since the plaquettes connected by the red line in the $z$ direction share the same holonomy, the cube is not excited. Hence, the energy of the state is reduced by two unities. However, it is still an excited state of the Toric Code. The ground state degeneracy of the Toric Code does not change under this modification of the Hamiltonian.  

By continuously extending the red line of figure \ref{fig:TC_excited2} to the boundary, we obtain figure \ref{fig:TC_excited3}. The red line closes into a non-contractible curve. The energy of the state is reduced to be equal to the length of the torus in the $z$ direction. The resulting state of figure \ref{fig:TC_excited3} is still an excited state of the Toric Code, and the corresponding modified Hamiltonian remains equivalent to $H_{TC}$.

The same procedure can be performed to the neighbouring cubes that host the other half of the dual ribbon, as shown in figures \ref{fig:TC_excited4} and \ref{fig:TC_excited5}. However, the resulting state in figure \ref{fig:TC_excited5} is a ground state. Thus, the Hamiltonian obtained at the end of the process shown in figures \ref{fig:TC_excited2}-\ref{fig:TC_excited5} defines a new model, in which ground states are given by closed dual surfaces and surfaces bounded by the red non-contractible curves. Its ground state degeneracy is $GSD_{TC} + 2$, where $GSD_{TC}$ is the ground state degeneracy of the Toric Code. 

The procedure of connecting parallel plaquettes by red lines can be extended to the whole lattice at every direction. The resulting Hamiltonian is the one given by equation \eqref{eq:H-3D-Z2}, and it describes the fracton-like model. In this way, there is a continuous process in which we can go from the Toric Code Hamiltonian to equation \eqref{eq:H-3D-Z2}. In this process, TC excitations condense into ground states of the fracton-like model. However, note that only TC excitations given by non-contractible ribbons condense into fracton-like ground states. For instance, an TC excitation given by a contractible surface, as in figure \ref{fig:ee1-3d}, is a fracton excitation in the fracton-like model. 

\subsubsection{Fracton excitations}\label{sec:fracton-exc}

The elementary excited states $\ket{\phi} \in \mathcal{H}$ are such that either $A_v\ket{\phi} = 0$, for some vertex $v$, or $B_c^{(\mu)}\ket{\phi} = 0$, for some cube $c$ and direction $\mu$. A string of $\sigma^z$ operators, beginning at a vertex $v$ and ending at a vertex $v'$, creates excitations of $A_v$ and $A_{v'}$, also called charge excitations. Since $A_v$ is essentially the gauge transformation of the $3D$ toric code, the charge excitations of the model \eqref{eq:H-3D-Z2} are the same charge excitations of the $3D$ Toric Code, and they can move freely in the lattice without an energy cost.

Now, excited states of the cube operators \(B_c^{(\mu)}\) are called $\mu$-flux excitations and can be pictured as lying at the corners of membranes. This can be better understood using the graphical representation of states as in figure \ref{fig:11}. The simplest \(\mu\)-flux excited state is created by the action of a \(\sigma^x\) operator on a single site over a ground state of the model. Note that whether this operator acts on a link along the \(x, y\) or \(z\)-axis will result on certain combinations of \(x, y \) and \(z\)-fluxes. For instance, acting with \(\sigma_l^x\) on a ground state: 
\begin{align}
    \ket{\phi} = \sigma_{l}^x \ket{\psi_0},
\end{align}
where \(l\) is a \(x\)-like link, results on a state with pairs of \(y\) and \(z\)-fluxes at the boundaries of the membrane as depicted in figure \ref{fig:ee1-3d}.

These excitations have restricted mobility since their localization is associated to the corners of the membrane. For example, the state represented by figure \ref{fig:ee2-3d} shows that extending the membrane along the \(y\)-direction move pairs of \(\mu\)-fluxes. In general, moving these excitations correspond to extending the membrane without changing the number of corners. On the contrary, if the membrane is bent towards its orthogonal direction more excited states are created increasing the energy of the state, as shown in figure \ref{fig:ee3-3d}. Again, this is interpreted as an energy penalization to deformations of membranes that change their number of corners.

\section{Entanglement Entropy}\label{ententr}
In this section, we calculate the entanglement entropy $S_A$ of a sub-region $A$ of the lattice for the models presented in sections \ref{sec2} and \ref{sec3}. For both cases, we interpret the result as a relation between the entanglement entropy of $A$ and the ground state degeneracy of a restriction of the full corresponding model to the sub-region $A$.

\subsection{Entanglement entropy of the $2d$ fracton-like model}\label{ee2d}

\begin{figure}
    \centering
    \includegraphics[scale=0.07]{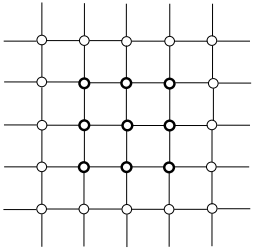}
    \caption{Division of the lattice into two sub-regions, $A$ and $B$, where region $A$ is composed by black vertices.}
    \label{fig:regionA-2d}
\end{figure}

Consider the $2d$ fracton-like model of section \ref{sec2}. The model is defined on a square lattice of size $L_x \times L_y$. Let's split the lattice into two sub-regions, $A$ and $B$, as in figure \ref{fig:regionA-2d}, where sub-region $A$ is characterized by black vertices and has size $R_x \times R_y$. The total Hilbert space $\mathcal{H}$ is thus given by the tensor product $\mathcal{H} = \mathcal{H}_A \otimes \mathcal{H}_B$, where $\mathcal{H}_A$ and $\mathcal{H}_B$ are the Hilbert spaces associated to the regions $A$ and $B$, respectively. We want to calculate the entanglement entropy $S_A$ of sub-region $A$. The density matrix of the model is given by

\begin{equation}\label{density_matrix}
    \rho = \frac{\Pi_0}{Tr(\Pi_0)} = \frac{1}{GSD}\Pi_0,
\end{equation}
where $GSD$ is the ground state degeneracy given by equation (\ref{eq4}) and $\Pi_0$ is the ground state projector, given by
\begin{equation}\label{ground_state_projector}
    \Pi_0 = A\prod_p B_p,
\end{equation}
where $A$ is given by equation (\ref{eq2}) and $B_p$ by equation (\ref{eq1}). The entanglement entropy is the \textit{von Neumann entropy} of the reduced density matrix
\begin{equation}\label{entanglemente_entropy}
    S_A = -Tr\left(\rho_A log( \rho_A)\right),
\end{equation}
where the reduced density matrix $\rho_A$ is obtained from $\rho$ by tracing out the region $B$, namely,
\begin{equation}\label{reduced_density_matrix}
    \rho_A = Tr_B(\rho).
\end{equation}
Taking the partial trace of $\rho$ over region $B$, we have
\begin{eqnarray}\label{reduced1}
    \rho_A  = \frac{1}{GSD}Tr_B\left[\frac{1}{2}\left(\bigotimes_{v\in K_0} \mathbb{1}_{v} + \bigotimes_{v\in K_0}\sigma^x_v\right)\prod_p B_p\right]
\end{eqnarray}
The second term of the sum in the right-hand side of equation (\ref{reduced1}) multiplies a $\sigma^x$ matrix to every vertex in the lattice. As we will see more clearly in what follows, the product $\prod_p B_p$ can be written as a sum over plaquettes of products of $\sigma^z$ matrices. In this sum, there will be terms where there will be no $\sigma^z$ matrix acting on the region $B$, and thus introducing a $\sigma^x$ operator to this region and taking the trace over $B$ will yield zero. For terms in the sum where there are $\sigma^z$ matrices acting on vertices in the region $B$, the multiplication by $\sigma^x$ results in $-i\sigma^y$, which also has trace zero. Therefore, the second term of the sum in the right-hand side of equation (\ref{reduced1}) is equal to zero, and we have
\begin{equation}\label{reduced3}
    \rho_A = \frac{1}{2GSD} Tr_B\left(\prod_p B_p\right).
\end{equation}
Now, using equation (\ref{eq1}), we can expand the product $\prod_p B_p$ into the sum given by equation (\ref{reduced4}):
\begin{widetext}
\begin{equation} \label{reduced4}
    \prod_p B_p = \prod_p\frac{1}{2}\left(\bigotimes_{v \in p}\mathbb{1}_v + \bigotimes_{v \in p}\sigma^z_v\right) = \frac{1}{2^{N_p}}\left(\bigotimes_{v \in K_0} \mathbb{1}_v + \sum_p \bigotimes_{v \in p}\sigma^z_v + \sum_{p \neq q}\bigotimes_{v \in p}\sigma^z_v\bigotimes_{v' \in q}\sigma^z_{v'} + ...\right),
\end{equation}
\end{widetext}
where $N_p$ is the number of plaquettes in the lattice. To each plaquette, we assign an operator of the form $b_p = \bigotimes_{v \in p}\sigma^z_v$ and a number $s_{p} \in \{0,1\}$. We also define the vector $s = (s_{p_1}, ..., s_{p_{N_p}})$, whose ith entry is the number $s_{p_i}$, associated to the ith plaquette in the lattice. Then, for each possible vector $s$, we define the product

\begin{equation}\label{reduced5}
    g_s = b_{p_1}^{s_{p_1}}b_{p_2}^{s_{p_2}}...b_{p_{N_p}}^{s_{p_{N_p}}}
\end{equation}
of operators $b_p$ for every plaquette of the lattice. To illustrate, the vector $s_0 = (0,...,0)$ gives the product $g_{s_0} = \bigotimes_{v}\mathbb{1}_v$, while the vector $s_1 = (1,0,...,0)$ gives the product $g_{s_1} = \bigotimes_{v \in p_1}\sigma^z_v$. It is straightforward to check that $g_s$, for every vector $s$, forms a finite Abelian group, which we call $G$. Also, the sum over all elements of $G$ is equivalent to the sum in the expansion shown in equation (\ref{reduced4}). Therefore, since the trace will give zero whenever there are $\sigma^z$ operators acting over the region $B$ and the only surviving terms will be those with only identity operators acting over $B$, we can write
\begin{eqnarray}
    \rho_A = \frac{1}{2^{N_p + 1}GSD}Tr_B\left(\sum_{g_s \in G}g_s\right) \nonumber
\end{eqnarray}
\begin{eqnarray}\label{reduced6}
     = \frac{1}{2^{N_p + 1}GSD}dim(\mathcal{H}_B)\sum_{g^A_s \in G}g^A_s,
\end{eqnarray}
where $g^A_s$ are the elements in $G$ which have only identity operators acting over the sub-region $B$, and they form a subgroup, which we call $G^A$. This implies that
\begin{equation}\label{reduced7}
    \rho_A^2 = \frac{dim(\mathcal{H}_B)}{2^{N_p + 1}GSD}|G^A|\rho_A,
\end{equation}
where $|G^A|$ is the order of the subgroup $G^A$. Then, from equations (\ref{entanglemente_entropy}) and (\ref{reduced7}) it is easy to see that the entanglement entropy for sub-region $A$ is given by
\begin{equation}\label{entropy1:2d}
    S_A = log\left(\frac{2^{N_p + 1} GSD}{|G^A|dim(\mathcal{H}_B)}\right).
\end{equation}
For a lattice with dimensions $L_x \times L_y$ and a sub-region $A$ with dimensions $R_x \times R_y$, with $GSD$ given by equation (\ref{eq4}), the entanglement entropy for the sub-region $A$ is
\begin{equation}\label{entropy:2d}
    S_A = log(2^{R_x + R_y + 1}) = R_x + R_y + 1.
\end{equation}

One can notice that there is a similarity between the functional dependence of the entanglement entropy (\ref{entropy:2d}) and the logarithm of the ground state degeneracy of the same model defined only on the sub-lattice $A$. In fact, this result agrees with those in \cite{ibieta2020topological}, where the entanglement entropy for some sub-region $A$, for arbitrary topological models, was shown to be equal to the logarithm of the ground state degeneracy of the model restricted in a particular way to the sub-region $A$, namely
\begin{equation}\label{topol_ent_entr}
    S_A = log(GSD_{\tilde{A}}),
\end{equation}
where $GSD_{\tilde{A}}$ is the ground state of the model restricted to sub-region $A$ and without the gauge transformations on the boundary of $A$. To see that this is indeed the case, let's calculate the ground state degeneracy of the model given by equation (\ref{eq3}) restricted to some sub-region $A$ of dimensions $R_x \times R_y$ and without gauge transformations on the boundary $\partial A$ of $A$. Note that, since the transformations (\ref{eq2}) act globally, the only way to exclude such transformations from the boundary $\partial A$ is to exclude them from the whole lattice. Thus, the restricted model is given by
\begin{equation}\label{restricted}
    H_{\tilde{A}} = - \sum_{p \in A}B_p.
\end{equation}
We can construct ground states of this model just like we did in section \ref{sec2}, but now every ground state we had will be split into two different ones. For example, the two states which form the superposition $\ket{\psi_0}$ of figure \ref{fig:phys-states}(a), which is a ground state of the full model, will be ground states of the restricted model (\ref{restricted}). Thus, the model (\ref{restricted}) has two times the degeneracy of the full model, and we have that
\begin{equation}\label{gsd:restricted2d}
    GSD_{\tilde{A}} = 2^{R_x + R_y + 1}.
\end{equation}
We can immediately see that equations (\ref{entropy:2d}) and (\ref{gsd:restricted2d}) satisfy equation (\ref{topol_ent_entr}).

\subsection{Entanglement entropy of the $3d$ fracton-like model}

Now, consider the $3d$ fracton-like model of section \ref{sec3}, defined on a cubic lattice of size $L_x \times L_y \times L_z$. The calculation of the entanglement entropy for this model will follow almost exactly the previous one. We split the lattice into two sub-regions, $A$ and $B$, where sub-region $A$ has dimensions $R_x \times R_y \times R_z$. The total Hilbert space $\mathcal{H}$ will then be given by the tensor product $\mathcal{H} = \mathcal{H}_A \otimes \mathcal{H}_B$ of the Hilbert spaces associated to sub-regions $A$ and $B$.  The density matrix is given by equation (\ref{density_matrix}), where $GSD$ is given by equation (\ref{eq10}) and $\Pi_0$ will now be given by
\begin{equation}\label{projector3dmodel}
    \Pi_0 = \prod_v A_v \prod_{\mu, c}B_c^{(\mu)},
\end{equation}
where, for every vertex $v$, $A_v$ is given by equation (\ref{eq:Av-3d1}) and, for every cube $c$, $B_c^{(\mu)}$ is given by equations (\ref{eq:Bc-3dx}), (\ref{eq:Bc-3dy}) and (\ref{eq:Bc-3dz}), for $\mu = x, y, z$ respectively. From the definitions of the cube operators, it follows that for a fixed cube $c$,
\begin{eqnarray}
     B_c^{(x)}B_c^{(y)}B_c^{(z)} = \frac{1}{4}\left(\mathbb{1} + Z_c^{(x)} + Z_c^{(y)} + Z_c^{(z)}\right), 
\end{eqnarray}
where
\begin{eqnarray}\label{eq:Zc-3d}
    Z_c^{(x)} \;\vcenter{\hbox{\includegraphics[scale=0.57]{Bc-ket1.eps}}}\, = & \;\vcenter{\hbox{\includegraphics[scale=0.57]{Bc-ket2.eps}}}, \label{eq:Zc-3dx}
\end{eqnarray}
\begin{eqnarray}
    Z_c^{(y)} \;\vcenter{\hbox{\includegraphics[scale=0.57]{Bc-ket1.eps}}}\, = & \;\vcenter{\hbox{\includegraphics[scale=0.57]{Bc-ket3.eps}}}, \label{eq:Zc-3dy}
\end{eqnarray}
\begin{eqnarray}
    Z_c^{(z)} \;\vcenter{\hbox{\includegraphics[scale=0.57]{Bc-ket1.eps}}}\, = & \;\vcenter{\hbox{\includegraphics[scale=0.57]{Bc-ket4.eps}}}. \label{eq:Zc-3dz}
\end{eqnarray}
Then, we have 
\begin{eqnarray}\label{project_exp}
    \Pi_0 = \frac{1}{2^{N_v + 2N_c}}\prod_v \left(\mathbb{1}_v + X_v\right)\prod_c (\mathbb{1}_c + \sum_{\mu}Z^{(\mu)}_c),
\end{eqnarray}
where $\mathbb{1}_v = \bigotimes_{l \in star(v)}\mathbb{1}_l$, $X_v = \bigotimes_{l \in star(v)}\sigma^x_{l}$, $\mu = x, y, z$, $N_v$ is the number of vertices and $N_c$ is the number of cubes in the lattice. This product is equal to a sum of products of Pauli matrices, and thus we can proceed just as we did in the case of the $2d$ model by defining a group $G$ whose elements $g_s$, indexed by a vector $s$ with $N_v + 2N_c$ components, with each component being 0 or 1, are given by products of Pauli matrices, in a similar way as in equation (\ref{reduced5}). The result is that
\begin{equation}
    \Pi_0 = \frac{1}{2^{N_v + 2N_{c}}}\sum_{g_s \in G}g_s.
\end{equation}
From equation (\ref{reduced_density_matrix}), the reduced density matrix $\rho_A$ is then given by
\begin{equation}
    \rho_A = \frac{1}{2^{N_v + 2N_c}}dim(\mathcal{H}_B)\sum_{g^{\tilde{A}}_s \in G}g_s^{\tilde{A}},
\end{equation}
where $\tilde{A}$ is a region obtained from $A$ by excluding the vertices at the boundary $\partial A$ of $A$. The reason this is the only surviving region after we take the trace over $B$ is that gauge transformations $A_v$ that act over vertices $v$ on the boundary of $A$ can introduce $\sigma^x$ operators acting over links outside of $A$, i.e., links that belong to $B$. The trace over $B$ then gives zero in such case. 
\par It is straightforward to check that the elements $g^{\tilde{A}}_s$ form a subgroup of $G$. Thus, we can perform exactly the same steps we did in section \ref{ee2d} and then, using equation (\ref{entanglemente_entropy}), it follows that the entanglement entropy of sub-region $A$ is given by
\begin{equation}
    S_A = log\left(\frac{2^{N_v + 2N_c}GSD}{|G^{\tilde{A}}|dim(\mathcal{H}_B)}\right),
\end{equation}
where $|G^{\tilde{A}}|$ is the order of the subgroup $G^{\tilde{A}}$ formed by the elements $g_s^{\tilde{A}}$. Since the lattice has dimension $L_x \times L_y \times L_z$ and region $A$ has dimension $R_x \times R_y \times R_z$, we have that
\begin{eqnarray}\label{entanglement3d}
    S_A = 3(R_xR_y + R_xR_z + R_yR_z) + 2,
\end{eqnarray}
and this result is also consistent with equation (\ref{topol_ent_entr}). To see this, let's calculate the ground state $GSD_{\tilde{A}}$ of the reduced model. To reduce the model, we must discard the gauge transformations on the boundary of region $A$. This means that configurations which differ from each other only by a gauge transformation on the boundary $\partial A$ are not gauge-equivalent anymore, and must be accounted for in the calculation of $GSD_{\tilde{A}}$. So, the number of configurations to be added to $GSD$ is the number of gauge transformations on the boundary $\partial A$, which is simply the number of vertices in $\partial A$, and this number is equal to $2(R_xR_y + R_xR_z + R_yR_z) + 2$. Thus, $GSD_{\tilde{A}} = R_xR_y + R_xR_z + R_yR_z + 2(R_xR_y + R_xR_z + R_yR_z) + 2$ and equation (\ref{topol_ent_entr}) holds.

The constant term in equation \eqref{entanglement3d} is equal to the topological entanglement entropy of the $3d$ Toric Code. This is expected because, as we discussed in section \ref{sec:gsd3d}, Toric Code ground states belong to the set of ground states of the fracton-like model.

\section{Remarks on generalizations to arbitrary gauge groups}\label{sec4}

The models presented in sections \ref{sec2} and \ref{sec3} can be generalized to models based on arbitrary, possibly non-Abelian finite groups. In this section, we make some remarks about how this generalization could be done. We leave a more in-depth discussion for future work.

\subsection{\(G\)-fractonlike order in two dimensions}\label{sub4.1}

Consider a $2$-dimensional oriented manifold $M$, discretized by a square lattice. At each vertex $v$, we have a local Hilbert space $\mathcal{H}_v$ with basis given by states $\ket{g}$ labelled by some element $g\in G$, where $G$ is an arbitrary finite group and the total Hilbert space is $\mathcal{H} = \bigotimes_v \mathcal{H}_v$. Global gauge transformations are given by
\begin{equation}\label{eq12}
    A^g|a,b,c,...\rangle = \ket{ga,gb,gc,...},
\end{equation}
where $g \in G$ and $\ket{a,b,c,...} \in \mathcal{H}$ is an arbitrary basis state in the total Hilbert space, with $a,b,c, ... \in G$. We define the operator $A$ as the normalized sum of all global gauge transformations, namely,  

\begin{equation}\label{eq11}
    A = \frac{1}{|G|}\sum_{g \in G}A^g.
\end{equation}
One can easily see that if $G = \mathbb{Z}_2$, we recover equation (\ref{eq2}). Next, we define two plaquette operators $B_p^{(\mu)}$, $\mu = x,y$, which in the language of higher gauge theories (see \cite{ricardo}),  act by comparing the $0$-holonomy of links that are parallel to the $\mu$ direction. They are given by the formulas
\begin{align}
   B_p^{(x)} \;\vcenter{\hbox{\includegraphics[scale=0.57]{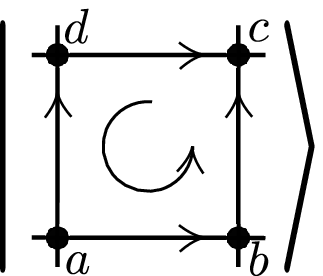}}}\, & = \delta (ab^{-1},dc^{-1}) \;\vcenter{\hbox{\includegraphics[scale=0.57]{Bp-ket.eps}}}\,, \label{eq:Bp-x-2d}
\end{align}
\begin{align}
   B_p^{(y)} \;\vcenter{\hbox{\includegraphics[scale=0.57]{Bp-ket.eps}}}\, & = \delta (ad^{-1},bc^{-1}) \;\vcenter{\hbox{\includegraphics[scale=0.57]{Bp-ket.eps}}}\,. \label{eq:Bp-y-2d}
\end{align}

If $G$ is Abelian, the two operators are actually the same, and one can easily check that if $G = \mathbb{Z}_2$, we recover equation (\ref{eq1}). The Hamiltonian is then given by

\begin{align}\label{eq13}
    H = -A - \sum_p\left(B_p^{(x)}+B_p^{(y)}\right).
\end{align}
\par One may notice a similarity between this theory and the usual Quantum Double Model of the finite group $G$ \cite{kitaev2, hbombin, buerschaper,ferreira20142d}. In fact, the operators defined in \eqref{eq12}, \eqref{eq:Bp-x-2d} and \eqref{eq:Bp-y-2d} satisfy the quantum double algebra $D(G)$. Define the local operators $L^g_v, R_v^g, T_v^g:\mathcal{H}_v \to \mathcal{H}_v$ such that, $\forall \ket{a} \in \mathcal{H}_v$,
\begin{align}
    L_v^g\ket{a} = \ket{ga}, \\
    R_v^g\ket{a} = \ket{ag}, \\
    T_v^g\ket{a} = \delta(g,a)\ket{a}.
\end{align}
These operators satisfy the quantum double algebra $D(G)$ \cite{kitaev2}. We can write $A$ and $B^{(\mu)}_p$, for every $p$, in terms of $L^g_v$, $R^g_v$ and $T^g_v$ in the following way: first, from \eqref{eq12}, we have that
\begin{align}\label{eq:a_qd}
    A^g = \bigotimes_v L^g_v.
\end{align}
Now, for every plaquette $p = (v_1v_2v_3v_4)$, we define the operators
\begin{align}
    B_p^{(x)}(g) = \sum_{\{b_i\}_{i=1}^4}\delta(b_1b_2^{-1}b_3b_4^{-1},g)\bigotimes_{v_i \in p}T^{b_i}_{v_i}, \label{bx_qd}\\
    B_p^{(y)}(g) = \sum_{\{b_i\}_{i=1}^4}\delta(b_1b_4^{-1}b_3b_2^{-1},g)\bigotimes_{v_i \in p}T^{b_i}_{v_i}. \label{by_qd}
\end{align}
For $g = e$, the group identity, $B_p^{(x)}(e) = B_p^{(x)}$ and $B_p^{(y)}(e) = B_p^{(y)}$, where $B_p^{(x)}$ and $B_p^{(y)}$ are defined in \eqref{eq:Bp-x-2d} and \eqref{eq:Bp-y-2d}, respectively. With these definitions, it is straightforward to check that the operators $A$ and $B_p$ do indeed satisfy the quantum double algebra of $G$, i.e., $\forall g, h \in G$,
\begin{align}
    A^gA^h = A^{gh}, \\
    A^gB^{(\mu)}_p(h) = B^{(\mu)}_p(ghg^{-1})A^g,
\end{align}
for any $\mu = x,y$. 

The algebra of the operators \eqref{eq11}, \eqref{eq:Bp-x-2d} and \eqref{eq:Bp-y-2d} is the quantum double of $G$, but the model \eqref{eq13} is different from the usual Quantum Double Models. Here, we have a quantum double algebra for each plaquette in the lattice, while in the usual QDM's, there is an algebra for each vertex-plaquette pair. A consequence of this fact is that there are no dyons here, only flux quasi-particles. However, the fact that the operator algebra in this model is the quantum double of $G$ allows us to immediately classify the flux quasi-particles of the model. A more in-depth discussion about this topic will appear in a future work.

\subsection{$G$-fractonlike order in three dimensions}\label{sec:G-fracton-3D}

Consider a $3$-dimensional oriented manifold $M$, discretized by a regular cubic lattice. At each link $l$, there is a local Hilbert space $\mathcal{H}_l$, generated by a set $\{\ket{g}\}$ of basis elements, labelled by some finite group $G$. The total Hilbert space is given by the product $\mathcal{H} = \bigotimes_l\mathcal{H}_l$.
We define three cube operators $B_c^{(\mu)}$, one for each direction $\mu = x, y, z$. $B_c^{(\mu)}$ compares the $1$-holonomy of plaquettes that are orthogonal to the $\mu$ direction, in the follwing way:
\begin{widetext}
\begin{align}
    B_c^{(x)} \;\vcenter{\hbox{\includegraphics[scale=0.98]{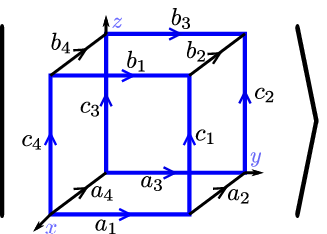}}}\, &= \delta (a_1 c_1 b_1^{-1} c_4^{-1}, a_3c_2b_3^{-1}c_3^{-1}) \;\vcenter{\hbox{\includegraphics[scale=0.98]{Bc-x.eps}}}\,, \label{eq:Bc-x}
\end{align}
\begin{align}
    B_c^{(y)} \;\vcenter{\hbox{\includegraphics[scale=0.98]{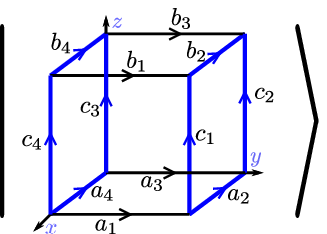}}}\, &= \delta (a_2 c_2 b_2^{-1} c_1^{-1}, a_4c_4b_4^{-1}c_4^{-1}) \;\vcenter{\hbox{\includegraphics[scale=0.98]{Bc-y.eps}}}\,, \label{eq:Bc-y}
\end{align}
\begin{align}
    B_c^{(z)} \;\vcenter{\hbox{\includegraphics[scale=0.98]{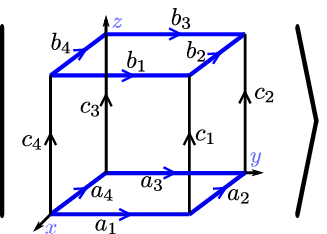}}}\, &= \delta (a_1 a_2 a_3^{-1} a_4^{-1}, b_1b_2b_3^{-1}b_4^{-1}) \;\vcenter{\hbox{\includegraphics[scale=0.98]{Bc-z.eps}}}\,.\label{eq:Bc-z}
\end{align}
\end{widetext}

It turns out that, for arbitrary groups, it is not possible to define an operator analogous to \eqref{eq:Av-3d1} which commutes with every cube operator. However, we can define a global operator $A$, inspired by the $2d$ model of section \ref{sub4.1}, made of a suitable combination of $L^g$'s and $R^g$'s to make it commute with all the cube operators. Then, we define a global operator
\begin{align}\label{eq:Av-g}
    A = \frac{1}{|G|}\sum_{g \in G}A^g,
\end{align}
where $A^g$ is defined as
\begin{align}
    A^g = \bigotimes_{p} L^g_{l_1}\otimes R^g_{l_2}\otimes R^g_{l_3}\otimes L^g_{l_4},  
\end{align}
where $l_i \in \partial p$, $i = 1,...,4$ and the plaquettes are oriented outwards.   

The Hamiltonian is defined as
\begin{equation}\label{eq19}
    H = -A - \sum_c\left(B_c^{(x)} + B_c^{(y)} + B_c^{(z)}\right),
\end{equation}
The operators defined in equation (\ref{eq:Av-g}), (\ref{eq:Bc-x}), (\ref{eq:Bc-y}) and (\ref{eq:Bc-z}) commute for every cube in the lattice, so the Hamiltonian (\ref{eq19}) can be diagonalized. Moreover, as in section \ref{sub4.1}, we can write these operators in terms of $L^g$'s, $R^g$'s and $T^g$, and then it can be shown that they also satisfy the quantum double algebra of $G$. We leave this discussion for a future work.

\section{Conclusions and outlook}
In this work, we have studied models of fracton-like order based on lattice gauge theory with subsystem symmetries in two and three spatial dimensions. The $3$-dimensional model reduces to the $3d$ Toric Code when subsystem symmetry is explicitly broken, realizing an example of a subsystem-symmetry enriched topological phase. It exhibits some features that are usually not present in the most common realizations of fracton order, such as a ground state degeneracy that depends exponentially on square of the linear size of the system and on its topology, and fully mobile excitations living along with fractons. However, the fracton-like character is destroyed if local perturbations that break the subsystem symmetry are applied to the system. We also calculated the entanglement entropy of these models, and we showed that it obeys a simple formula which was derived for the case of usual topological models, relating the entanglement entropy and the ground state degeneracy of a particularly restricted model. Although they are not topologically protected, these models show that one can obtain fracton-like phases from more regular lattice gauge theories, giving an alternative to the usual constructions found in the literature. 

One important open question is how the models introduced in this work and more standard models such as the X-cube model are connected. Since the ground state degeneracy of the model described in section \ref{sec3} scales differently from the X-cube, a relation between the two is not obvious. Likewise, it is not clear what is the relation between the generalized lattice gauge theory with subsystem symmetries of \cite{Vijay16, slagle1, williamson2016fractal} and the approach developed in this work. 

Moreover, the definitions of the models in sections \ref{sec2}, \ref{sec3} and \ref{sec4} show an explicit dependence on the geometry and the topology of the system, which suggests that new phenomena may arise if we define the models in manifolds with non-trivial geometry, topology and discretization. This direction was pursued for the X-cube model \cite{slagle2018x,shirley2018fracton}, and therefore it may also be helpful in the quest to clarify how the two models are connected. 

At the end of this work we made some remarks about the possibility of constructing fracton-like models from non-Abelian gauge groups. We argued that the operator algebra of these models is the quantum double of the corresponding finite group, but the models constructed in this way are not equivalent to the known Quantum Double Models. This direction is worth pursuing because this construction allows us to study more directly the behavior of non-Abelian fractons, which are known in the literature \cite{prem2019cage,song2019twisted}. This could possibly lead to a better understanding on how to apply fracton phases in quantum computation. On the other hand, the entanglement entropy can certainly give more information about the nature of entanglement in the ground/excited states of the fracton models we introduce in this work. In \cite{ibieta2020topological}, we show that the entanglement entropy calculation can be mapped into the counting of edge states in the entanglement cut. This also holds for the fracton-like models of this work and, if it is also the case for more standard fracton models is an immediate question worthy of further study that could deepen our understanding of gapped quantum phases of matter.

\acknowledgments
We thank Kevin Slagle, Dominic J. Williamson and Abhinav Prem for useful comments and suggestions. JPIJ thanks CNPq (Grant No. 162774/2015-0) for support during this work. LNQX thanks CNPq (Grant No. 164523/2018-9) for supporting this work. MP is supported by Capes.

\bibliographystyle{unsrt}
\bibliography{bib}

\end{document}